\documentclass[prl,twocolumn,showpacs]{revtex4}
\usepackage[dvips]{graphicx}          
\usepackage{amsfonts,amssymb,wasysym} 
\usepackage{psfrag}                   
\newcommand{\LoopCross}{$\Circle \!\! - \!\! \Circle$}

\begin{document}
\title{The dynamics of critical Kauffman networks under asynchronous stochastic update}
\author{Florian~Greil and Barbara~Drossel} 
\affiliation{Institut f\"ur Festk\"orperphysik,  TU Darmstadt,
Hochschulstra\ss e 6, 64289 Darmstadt, Germany }
\date{\today}
\begin{abstract}
We show that the mean number of attractors in a critical Boolean 
network under asynchronous stochastic update grows like a power law 
and that the mean size of the attractors increases as a stretched exponential
with the system size. This is in strong contrast to the synchronous
case, where the number of attractors grows faster than any power law.
\end{abstract} 
\maketitle

Random Boolean networks were introduced in 1969 by Kauffman
 \cite{kaufmann:metabolic, kaufmann:homeostasis} as a simple model for
 complex systems consisting of units that interact via directed
 links. They are used to model social and economic networks
 \cite{alexander:random, paczuski:self-organized}, neural networks,
 and gene or protein webs \cite{kauffman:random}.

A random Boolean network (RBN) is a directed graph with $N$~nodes each
of which takes a Boolean value~$\sigma_i \in \{ 0, 1 \}$.  The
number~$k$ of incoming edges is the same for all nodes, and the
starting points of the edges are chosen at random. Usually these
models are updated synchronously, $\sigma_i(t+1) = c_i
(\sigma_{d_{i,1}}(t), \ldots, \sigma_{d_{i,k}}(t) )$, where the
Boolean coupling function~$c_i$ of node $i$ is chosen at random among
a given set of functions and $d_{i,j}$ denotes the $j$-th input of node~$i$. 
The configuration of the system~$\vec{\sigma}
\equiv \{\sigma_1, \ldots, \sigma_N \}$ thus performs a trajectory in
configuration space. \emph{Critical} networks are of special interest
\cite{aldana-gonzalez:boolean}.  Their dynamics is at the boundary
between a frozen phase where initially similar configurations
converge, and a chaotic phase where initially similar configurations
diverge exponentially.

However, synchronous update is highly improbable in real
networks, and it is used under the tacit assumption
that going to asynchronous update will not modify the essential
properties of the system \cite{harvey:time}. 
But there are good reasons to doubt the
validity of this assumption. For instance, for cellular automata it is
well-known that some of the self-organization is an artifact of the
central clock \cite{ingerson:structure}. For RBNs there is also
recent evidence that deviations from synchronous update modify
considerably the attractors of the dynamics \cite{klemm:stable, klemm:robust}.

In this paper, we investigate a version of the model where at each
computational step one node is chosen at random and is updated.
A model with this asynchronous stochastic updating scheme is often
called Asynchronous RBN (ARBN)\cite{gershenson:updating,
gershenson:contextual, gershenson:phase}, while the
Classical synchronous RBN is referred to as CRBN.  ARBNs are mostly
studied numerically with the focus on various measures of stability
\cite{di-paolo:rhythmic, rohlfshagen:circular, hallinan:asynchronous, mesot:critical, matache:asynchronous}. ARBNs were observed to be capable to generate an ordered 
behavior, but the detailed properties of attractors have not been studied yet.
 
We will show mostly analytically that the number of attractors changes completely
when going from CRBNs to ARBNs.  For CRNBs, attractors are cycles in
configuration space, and their number was recently shown numerically
\cite{kaufmann:metabolic, bilke:stability, socolar:scaling, bastolla:modular, bastolla:relevant} and analytically
\cite{samuelsson:superpolynomial, drossel:number}
to grow faster than any power law with the network size~$N$. 
In contrast, we will show in the following that for
asynchronous stochastic update the mean number of attractors grows as
power law while their size increases like a stretched exponential
with~$N$.  As the dynamics is no longer deterministic, an appropriate
definition of an attractor must be given. An attractor is a subset of
the configuration space such that for every pair of configurations on
the attractor there exists a sequence of updates that leads from one
configuration to the other. In \cite{harvey:time}, such an attractor
is called ``loose attractor''. Starting from a random initial
configuration, the system will eventually end up on an attractor.

Let us first consider a set of nodes arranged in a loop. Such loops
occur as relevant components of critical networks. Nontrivial dynamics
occurs only if the two constant Boolean functions are omitted, the
remaining Boolean functions being ``copy'' ($\oplus$), and ``invert''
($\ominus$). A loop with $n$ inversions $\ominus$ can be mapped
bijectively onto a loop with $n-2$ inversions by replacing two
$\ominus$ with two $\oplus$ and by inverting the state of all nodes
between these two couplings. It is therefore sufficient to consider
loops with zero inversions (``even'' loops) and loops with one
inversion (``odd'' loops). The position of the $\ominus$-coupling in
the odd loop is called the \emph{twisted edge}. For synchronous
update, each configuration is on a cycle in configuration space and
occurs again at most after $N$ (2$N$) time steps for even (odd)
loops. The number of cycles increases therefore exponentially with
$N$.  In contrast, most configurations are transient in the
asynchronous case, and only two (one) attractors are left. The reason
for this is that a domain of neighboring nodes that have the same
value increases or decreases with probability $1/N$ per computational
step. The domain size therefore performs a random walk, and for an
even loop no domain is left after of the order of $N^3$ updates. The
attractors are the two fixed points. For an odd loop, the
nodes of a domain change their state at the twisted edge, and the total 
number of domain walls is therefore odd. The attractor contains only 
one domain wall that moves around the loop, and the attractor comprises 
$2N$ configurations. The dynamics on such a loop is closely related to 
the Glauber dynamics~\cite{glauber:time-dependent} of a one-dimensional
Ising chain with cyclic boundary condition at temperature~$T=0$, where
the domains also shrink and grow with a fixed rate and where the
equal-time correlation function~\index{correlation function} obeys a
scaling form $C(r,t) = f(r^2 t^{-1})$~\cite{bray:universal}.  The
dynamics of an odd loop can be mapped onto the dynamics of an Ising
chain with one negative coupling. It is a frustrated system in which
not all bonds can be satisfied simultaneously. To conclude, by going
from synchronous to asynchronous update, the number of attractors of a
loop is reduced from an exponentially large number to 1 or 2. This was
also pointed out in \cite{klemm:topology}, where a different
asynchronous updating rule is used.

Let us next consider critical networks with connectivity $k=1$, where
the Boolean coupling functions are again ``copy'' and ``invert''.
\emph{Relevant nodes} are those nodes whose state is not constant and
that control at least one relevant element
\cite{bastolla:modular}. They determine the attractors of the
system. In \cite{flyvbjerg:exact}, exact results for the topology of
$k=1$ networks are derived, and in~\cite{drossel:number} it is shown
that the number of attractors of a critical $k=1$ network increases
faster than any power law.  The number of relevant nodes scales as
$\sqrt{N}$, and they are arranged in of the order of $\ln (N)$
loops. The remaining nodes sit on trees rooted in these loops. Under
asynchronous update, each loop has at most two attractors. The nodes
on trees rooted in even loops are frozen because the loop is on a
fixed point. The nodes on trees rooted in odd loops can assume any
combination of states, since one can find to each possible state of a
tree a sequence of updates that generates it. Since each loop is even
or odd with equal probability, the mean number of attractors of
networks with $n$ relevant loops is
\begin{eqnarray}
\frac 1 {2^n}\sum_i {n \choose i}2^i &=& \left(\frac{3}{2} \right)^n
\simeq \left(\frac{3}{2}\right)^{\ln (N/2)},
\end{eqnarray}
which is a power law in $N$. 
The number of nodes in trees is of the order of $N$. On average, half
of the trees are rooted in odd loops. Consequently the mean attractor
size increases exponentially with $N$. 

Finally, we investigate the most frequently studied critical networks
with connectivity $k=2$, where each of the 16 possible Boolean
coupling functions is chosen with equal probability. All nodes apart
from the order of $N^{2/3}$ nodes are frozen. This follows for
instance from the factor $\epsilon^3$ in the last term of Eqn.~(9) of
\cite{samuelsson:superpolynomial}, which implies that only the
fraction $N^{-1/3}$ of all nodes undergo a nonfrozen sequence of
states in time. Numerical support for this result is presented in
\cite{socolar:scaling}. 

The number of relevant nodes scales as $N^{1/3}$
\cite{socolar:scaling}, and only a fraction $N^{-1/3}$ of these
relevant nodes have two relevant inputs. This last statement follows
from the evaluation of Eqns.~(6) and (9) in
\cite{samuelsson:superpolynomial} in saddle-point approximation, where
the width of the first term in the direction perpendicular to the line
of maxima is of the order $N^{-2/3}$, implying that of the order
$N\cdot N^{-2/3} = N^{1/3}$ nodes have 2 nonfrozen
inputs. Consequently, the proportion $N^{-1/3}$ of nonfrozen nodes
(whether they are relevant or not) have 2 nonfrozen inputs. This
result, together with the other just mentioned features of the model
is confirmed by (unpublished) studies of our group.  The other
relevant nodes have one relevant input (as the second input comes from
a frozen node). The remaining non-frozen nodes (of the order of 
$N^{2/3}$) are on trees rooted in relevant nodes. 
Just as for the $k=1$ networks, there are of the order of $\ln (N)$ 
independent relevant components \cite{bastolla:modular}.
In contrast to the $k=1$ networks, these components are not always
simple loops, but may contain several nodes with two relevant
inputs. In order to obtain results for the number of attractors of the
networks, we have to investigate the attractors of such relevant
components.

Let us first consider relevant components that contain one node with
two relevant inputs. These are two loops with one interconnection
(\LoopCross-component), and a loop with an extra link
($\oslash$-component). The dynamics under synchronous update for such
components is studied in \cite{kaufman:on}, and it is found that the
number of attractors in both systems increases exponentially with the
number of nodes. With asynchronous update, the number of attractors
becomes very small. 

We discuss first two loops with one
interconnection. The first loop is independent of the second loop, and
its attractor is either a fixed point (if the loop is even), or it has
one domain wall moving around the loop. If the first loop is on a
fixed point, it provides a constant input to the second loop, which
therefore behaves like an even loop, or an odd loop, or a frozen
loop. The system can have at most three attractors. If the first loop
is odd, if provides a changing input to the second loop, which can
therefore have an attractor that contains an arbitrary and fluctuating
number of domain walls. Consequently, a loop that has one external
input can show one out of four different types of behavior on an attractor: 
\begin{enumerate}
\item The loop can be at a fixed point $0$.
\item The loop can be at a fixed point $1$.
\item There is exactly one domain wall which moves around the loop.
\item The number of domain walls in the loop fluctuates.
\end{enumerate}
(Without loss of generality, we have assumed that all coupling
functions for nodes with one input are ``copy''.)

\begin{figure}[htb]
\psfrag{a}{a}\psfrag{b}{b}\psfrag{c}{c}\psfrag{d}{d}\psfrag{e}{e}
\psfrag{f}{f}\psfrag{g}{g}\psfrag{h}{h}\psfrag{i}{i}
\includegraphics[width=0.45\textwidth]{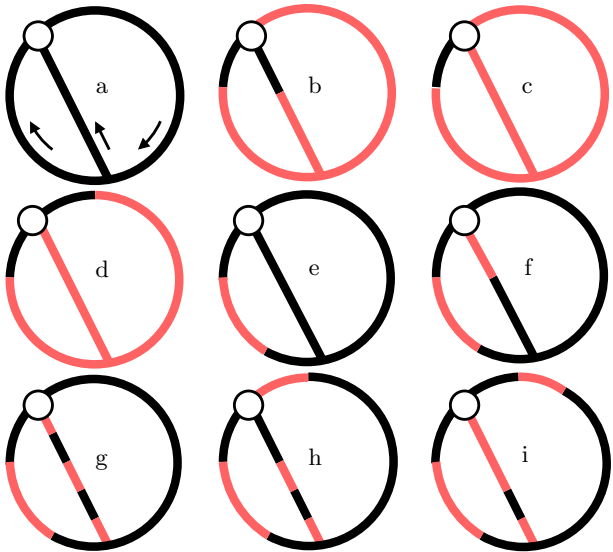}
\caption{A possible sequence of configurations of a loop with a cross-link
showing how multiple domain walls are generated. 
The coupling function of the node~$\Circle$
with two inputs is such that only two dark inputs lead to light output,
all other combination give black output. After (e) the procedure described by
(a) to (d) is repeated to obtain (f) and similarly for (g), (h) and (i).}
\label{CrossDevelop}
\end{figure}

Now we turn to a loop with an extra link. We can again assume that all
coupling functions for nodes with one input are ``copy''. If this
component has one fixed point (two fixed points), it is (they are) the
only attractor(s). This is because one can reach a fixed point from an
arbitrary initial state by updating one node after another by going
around the loop in the direction of the links. After at most two
rounds the fixed point is reached. Only if the coupling function for
the node with two inputs has no fixed point, a more complicated
attractor occurs. Without loss of generality, we choose this function
to have the output 1 if and only if both inputs are 0. By considering
the possible update sequences, one finds that the component can
accumulate a large and fluctuating number of domain walls, as illustrated 
in Fig.~\ref{CrossDevelop}.

Equipped with these results, we now consider components with several
nodes with 2 inputs. We define a section to be a sequence of nodes
starting at a node with two inputs and ending right before a node with
2 inputs. Such a sequence can branch and have several end
points. Clearly, the number of sections is the number of nodes with 2
inputs; a simple loop is counted as one section. A section is controlled 
by its first node, which is the one with 2 inputs. 
Just as for the loop with one external input, a section can
show on an attractor one out of the four different types of behavior
listed above. This is because all states that have more than a single
domain wall in a given section must be part of the same attractor. We show
this by the following argument: Assume that on an attractor there
occur two domain walls in a section.  The two domain walls can be
destroyed by updating all nodes between the two walls, such that the
domain enclosed by the walls vanishes.  A configuration with no wall
on the section (and with the state of all other sections unmodified) is
therefore also part of the attractor, and there exists consequently a
way back to the configuration with two domain walls on this
section. By repeating the same sequence of updates, every even number
of domain walls can be created in this section, and odd numbers can be
created by moving one domain wall out of the section. If $s$ is the
number of sections, an upper bound for the number of attractors of the
component is therefore given by $4^s$. 

We checked this analytical result by computer simulations. In order to
make sure that we capture all attractors, we did a complete search of
state space, which can only be done for small networks. Starting from
an initial state, we did $N^3$ updates before assuming that the
system is on an attractor, and we made sure that the results are not
changed when the length of the initial time period is varied. All
states that can be reached from this last state are on the same
attractor as this state. All other states that have been visited are
marked as transient states. Then we start with an unvisited state as
new initial condition in order to identify further transient states
and attractors. We constructed relevant components by starting with
one loop of a certain size, and by iteratively inserting additional
connections between two randomly chosen nodes.  The new connection
contains a randomly chosen number of 1 to 4 nodes (such that a section
can contain two domain walls in its interior). In these networks, the
number of sections, $s$, is identical to the number of nodes with 2
inputs, $\mu$. After each insertion, we evaluated the number of attractors
for different choices of coupling functions. This procedure was
repeated more than 750\,000~times. The largest number of attractors found in a
system is shown in Table~\ref{SimTab} as function of $s = \max(1,\mu)$.

\begin{table}[htb]
\begin{tabular}{ccrr}
\hline 
$\mu$ & $\nu_{\rm max}$ &$4^s$ & realizations \\
\hline \hline
0 & 2  &   1  & 227\,683\\
1 & 2  &   4  & 167\,370\\
2 & 9  &  16  & 138\,541\\
3 & 8  &  64  & 110\,263\\
4 & 23 &  256 & 73\,268 \\
5 & 25 & 1\,024 & 40\,770 \\
6 & 23 & 4\,096 & 15\,727 \\
\hline
\end{tabular}
\caption{Maximum number of attractors $\nu_{\rm max}$ as function of
the number of nodes with 2 inputs, $\mu$, for networks with up to
17~nodes. Networks with higher $\mu$ are probed less often because 
if only short links are added there is no node left which has not
already two inputs.}
\label{SimTab}
\end{table}

This leads us to the conclusion that a network consisting of the order
of $\ln(N)$ relevant components, with component number $i$ having
$\mu_i$ nodes with 2 inputs, cannot have more than 
\begin{eqnarray}
\nu = 4^{\max (1, \mu_1)} \! \cdot 4^{\max (1,\mu_2)} 
\! \cdot \ldots \cdot 4^{\max (1,\mu_{\ln N})} \!\! \leq 4^{\ln (N) + \mu}
\end{eqnarray}
attractors. This is a power law in $N$ if the probability distribution
for the value of $\mu$ becomes independent of $N$ for
large~$N$. Indeed, as we have mentioned above,  each of the $N^{1/3}$
relevant nodes has two (randomly chosen) relevant inputs with
probability $aN^{-1/3}$ (with some constant $a$). Since this
probability is independent for different nodes, the value of $\mu$ is
distributed for large $N$ according to a Poisson distribution with a
mean $a$.

We thus have shown that in critical $k=2$ networks with
asynchronous stochastic update, the number of attractors grows as a
power law in~$N$, which is in strong contrast to the synchronous case,
where the number of attractors increases like a stretched exponential
function.

We conclude with a discussion of the size of attractors in these
networks.  There are of the order of $N^{2/3}$ nodes on the trees
rooted in the relevant components.  These nodes in trees can adopt any
configuration if the node they are rooted in can switch its state on
an attractor. Since a non-vanishing fraction of all relevant nodes switch their states on an attractor,  the size of the
attractor is of the order of 
\begin{eqnarray}
 2^{N^{2/3}} = \exp \left( N^{2/3} \ln 2 \right)\, .
\end{eqnarray}
The size of the attractors grows like a stretched exponential function and
therefore faster than any power law. 

Many of our results hold also for other kinds of stochastic
asynchronous update, for instance if a certain (small) fraction of
nodes is updated at each step, or if the time interval between two
updates of a node is peaked at a value $\tau$ and Gaussian distributed
around it. (The latter case describes our system for large $N$, when
the network of relevant nodes is coarse-grained such that of the order
of $N^{1/3}$ neighboring nodes are replaced by a single node that
receives a delayed input from the previous node.) In these modified
stochastic models, domain walls on an isolated loop can annihilate,
but cannot be created again, leading to the same attractors as with
the completely stochastic update. However, the state of the trees
rooted in the loops will be dominated by a few domain walls when the
distribution of update times becomes narrow, with states with more
domain walls occurring rarely. Similarly, relevant loops that receive
input from outside, and relevant components with nodes with two inputs
will have attractors dominated by few domain walls, and the actual
size of attractors becomes in the thermodynamic limit $N\to \infty$
smaller than the size obtained by considering any possible sequence of
updates.

The biological implications of these findings are clear and have been
pointed out in \cite{klemm:topology}. Since biological networks do not
have a completely synchronous update, the number of attractors should
be derived from models with asynchronous update. 
In \cite{klemm:stable} it is found numerically that the number of stable
attractors increases sublinearly. Attractors are called stable if they 
do not change when a perturbation is added to a synchronous updating rule.
The present paper extends this finding by showing that the number of 
attractors in critical asynchronous Kauffman models increases as a power 
law, and we thus regain the original claim by Kauffman - albeit for models
with a different update rule than the original one.

\begin{acknowledgments}
\noindent We thank V.~Kaufman for useful discussions. 
\end{acknowledgments}

\bibliography{ArbnBib.bib}
\end{document}